\documentclass[longabstract,manuscript,dvips]{aastex}
\usepackage{graphicx}
\usepackage[latin1]{inputenc}
\usepackage{amssymb}
\usepackage{xspace}

\newcommand{\tsr}{{TSR}\xspace}
\newcommand{\eminus}{{\rm e}^{-}}
\newcommand{\kt}{k_{B}T}
\newcommand{\ktpar}{\kt_{\parallel}}
\newcommand{\ktperp}{\kt_{\hspace*{-0.15em}{\perp}}}

\renewcommand{\figurename}{Fig.}

\newcommand{\fefourteenplus}{\mbox{\ion{Fe}{15}}\xspace}
\newcommand{\fethirteenplus}{\mbox{\ion{Fe}{14}}\xspace}
\newcommand{\fetenplus}{\mbox{\ion{Fe}{11}}\xspace}
\newcommand{\fetenplusc}{\mbox{Fe$^{10+}$}\xspace}
\newcommand{\fenineplus}{\mbox{\ion{Fe}{10}}\xspace}
\newcommand{\fenineplusc}{\mbox{Fe$^{9+}$}\xspace}
\newcommand{\feeightplus}{\mbox{\ion{Fe}{9}}\xspace}
\newcommand{\feeightplusc}{\mbox{Fe$^{8+}$}\xspace}
\newcommand{\fesevenplus}{\mbox{\ion{Fe}{8}}\xspace}

\title{Electron-ion Recombination
	of \fenineplus forming \feeightplus
	and of \fetenplus forming \fenineplus: Laboratory 
	Measurements and Theoretical Calculations}

\author{M.~Lestinsky\altaffilmark{1,2}, 
	N.~R.~Badnell\altaffilmark{3},
	D.~Bernhardt\altaffilmark{4},
	M.~Grieser\altaffilmark{2},
	J.~Hoffmann\altaffilmark{2},
	D.~Luki\'c\altaffilmark{1,5},
	A.~M\"uller\altaffilmark{3},
	D.~A.~Orlov\altaffilmark{2}, 
	R.~Repnow\altaffilmark{2},
	D.~W.~Savin\altaffilmark{1},
	E.~W.~Schmidt\altaffilmark{4},
	M.~Schnell\altaffilmark{2},
	S.~Schippers\altaffilmark{4},
	A.~Wolf\altaffilmark{2},
	and
	D.~Yu\altaffilmark{4,6}
	}

\date{\today}
\email{lestinsky@astro.columbia.edu}

\altaffiltext{1}{Columbia Astrophysics Laboratory, Columbia University, New York, NY 10027, USA}
\altaffiltext{2}{Max-Planck-Institute for Nuclear Physics, 69117 Heidelberg, Germany}
\altaffiltext{3}{Department of Physics, University of Strathclyde, Glasgow G4 0NG, UK}
\altaffiltext{4}{Institut f\"ur Atom- und Molek\"ulphysik, Justus-Liebig-Universit\"at Giessen, D-35392 Giessen, Germany}
\altaffiltext{5}{Present address: Institute of Physics, Belgrade, 11080 Zemun, Serbia}
\altaffiltext{6}{Permanent address: Institute of Modern Physics, Chinese 
Academy of Sciences, Lanzhou 730000, P. R. China}
\keywords{atomic data --- atomic processes --- galaxies: active --- galaxies: nuclei --- plasmas --- X-rays: galaxies}

\slugcomment{This is an author-created, un-copyedited version of an 
article accepted for publication in Astrophysical Journal. IOP 
Publishing Ltd is not responsible for any errors or omissions in this 
version of the manuscript or any version derived from it.}

\begin{abstract}
We have measured electron-ion recombination for \fenineplusc forming
\feeightplusc and for \fetenplusc forming \fenineplusc using a merged beams
arrangement at the \tsr heavy-ion storage-ring in Heidelberg, Germany.
The measured merged beams recombination rate coefficients (MBRRC) for
relative energies from $0$ to $75$~eV are presented, covering all
dielectronic recombination (DR) resonances associated with
$3s\rightarrow3p$ and $3p\rightarrow3d$ core transitions in the 
spectroscopic species \fenineplus and \fetenplus, respectively.
We compare our experimental results to state-of-the-art 
multi-configuration Breit-Pauli (MCBP) calculations  and find significant 
differences.
Poor agreement between the measured and theoretical resonance structure is
seen for collision energies below $48$~eV for \fenineplus and below $35$~eV for
\fetenplus.  The integrated resonance strengths, though, are in
reasonable agreement.  At higher energies, good agreement is seen for
the resonance structure but for the resonance strenghts theory is 
significantly larger than experiment by a factor of $\approx 1.5$ (2) for 
\fenineplus (\fetenplus).
From the measured MBRRC we have extracted the DR contributions and
transform them into plasma recombination rate coefficients (PRRC) for
astrophysical plasmas with temperatures in the range of $10^2$ to
$10^7$~K.  This range spans across the regimes where each ion forms in
photoionized or in collisionally ionized plasmas.  For both temperature
regimes the experimental uncertainties are $25\%$ at a $90\%$ confidence
level.  As expected based on predictions from active galactic nuclei (AGN) observations as well
as our previous laboratory and theoretical work on M-shell iron, the
formerly recommended DR data severely underestimated the rate
coefficient at temperatures relevant for photoionized gas.  At these
temperatures relevant for photoionized gas, we find agreement between
our experimental results and MCBP theory. This is somewhat surprising
given the poor agreement in MBRRC resonance structure.
At the higher temperatures relevant for collisionally ionized gas, the MCBP 
calculations yield a \fetenplus DR rate coefficent which is significantly
larger than the experimentally derived one.
We present parameterized fits to our experimentally DR PRRC for ease of 
inclusion into astrophysical modelling codes.
\end{abstract}

\begin{document}
\maketitle

\section{Introduction}
\label{introduction}

Recent Chandra and XMM Newton X-ray observations of active galactic 
nuclei (AGNs) have detected
a new absorption feature in the 15-17 \AA\ wavelength range.  This has
been identified as an unresolved transition array (UTA) due mainly to
$2p \to 3d$ inner shell absorption in iron ions with an open M-shell
(Fe~{\sc i}-{\sc xvi}).  This spectral feature is believed to originate
in the warm absorber material surrounding the central supermassive
black hole in AGNs.  It has been identified in a number of
AGNs including
IRAS 13349+2438 \citep{Sako2001},
Mrk 509 \citep{Pounds2001},
NGC 3783 \citep{Blustin2002, Kaspi2002, Behar2003},
MCG -6-30-15 and Mrk 766 \citep{Sako2003},
NGC 5548 \citep{Steenbrugge2003},
I Zw 1 \citep{Gallo2004},
MR 2251-178 \citep{Kaspi2004},
Akn 564 \citep{Matsumoto2004},
NGC 4051 \citep{Pounds2004},
NGC 985 \citep{Krongold2005},
IC 4329A \citep{Steenbrugge2005}, and
NGC 3516 \citep{McKernan2007}.

Spectra from AGNs are typically analyzed using publicly available
photoionization codes such as Cloudy \citep{Ferland1998} and XSTAR
\citep{KallmanBautista2001, Kallman2004}. The modelled temperature and 
ionization balance of photoionized gas depends on the ionization parameter 
which is the ratio of the radiation field strength to the gas density.
A number of different definitions exist for this parameter.
Here we use $\xi = L/nR^2$ where $L$ is the luminosity of the radiation 
source, $n$ the gas density, and $R$ the distance to the source.

Based on atomic structure calculations and photoabsorption modeling,
\cite{Behar2001} have shown that the shapes, central wavelengths,
and equivalent widths of the UTAs can be used to diagnose the
properties of AGN warm absorbers.  However, models which match 
absorption features from second- and third-row elements cannot
reproduce correctly the observed UTAs from the fourth-row element
iron.  At the value of $\xi$ inferred from the second and third row
elements, the models appear to predict too high an ionization level
for iron, as was noted for NGC 3783 by \cite{Netzer2003}.  
They attributed this discrepancy to an underestimate of the low-temperature
dielectronic recombination (DR) rate coefficients for the Fe M-shell ions used in the models.  
To investigate this possibility \cite{Netzer2004} and \cite{Kraemer2004}
arbitrarily increased the low-temperature Fe M-shell DR rate
coefficients.  Their model results obtained with the modified DR rate
coefficients support the hypothesis of \cite{Netzer2003}.

These recent AGN observations have motivated experimental studies of
low-temperature DR for Fe M-shell ions \citep{Schmidt2006,
Schmidt2008, Lukic2007} and theoretical studies for Na- through Ar-like 
iron \citep{Gu2004, Badnell2006a, Badnell2006b, Altun2006, Altun2007}.
Here and throughout this paper the recombining systems are identified by 
their initial charge state.
In the temperature ranges where these ions form in photoionized plasmas 
these new Fe M-shell DR data are up to orders of magnitude larger than the 
DR data previously available. They are even significantly larger than the 
ad-hoc modified DR rate coefficients of \cite{Netzer2004} and \cite{Kraemer2004}.
These new Fe M-shell DR data appear to reduce the discrepancies involving 
the Fe M-shell UTA issues noted above.

Spectral fits to NGC 3783 using the DR data available before $\sim$~2006
yield multiphase models with a $\log(\xi) \sim 2$~erg~cm~s$^{-1}$
component to account for absorption features from second and third
row elements and a $\log(\xi) \sim 0$~erg~cm~s$^{-1}$ component to
account for the Fe M-shell UTA \citep[e.g.,][]{McKernan2007, Kallman2008}. 
Fits using the DR data available after $\sim$~2006 find little change
in $\log(\xi)$ derived from the second and third row elements.  
However, using these later DR data, the bulk of the Na- through Ar-like Fe ions 
form at a $\log(\xi)$ which is closer to the value inferred from 
absorption lines in the spectrum due to the second and third row elements
\citep{Kallman2008}.

DR also plays an important role in determining the thermal stability
of AGN warm absorbers.  It has long been known that such gas is thermally 
unstable at temperatures of $\sim 10^5$~K 
\citep{Krolik1981, ReynoldsFabian1995, Hess1997, Chakravorty2008}.
\cite{Hess1997} showed that Fe L-shell
ions play an important role in determining the range in parameter
space over which photoionized gas is predicted to be thermally
unstable.  This instability is robust to changes in elemental
abundances, shape of the ionizing spectrum, particle density, and
optical depth \citep{ReynoldsFabian1995, Hess1997}.  \cite{Savin1999} 
explored how uncertainties in the Fe L-shell ion DR data
available at that time affected the predicted instability and found
that varying the DR rate coefficients dramatically changed the range
in parameter space over which the gas was unstable.  Attempts to
understand this instability
motivated a series of Fe L-shell DR measurements for
F-like \ion{Fe}{18} \citep{Savin1997, Savin1999},
O-like \ion{Fe}{19} \citep{Savin1999, Savin2002a},
N-like \ion{Fe}{20} \citep{Savin2002b},
C-like \ion{Fe}{21} and
B-like \ion{Fe}{22} \citep{Savin2003}, and
Be-like \ion{Fe}{23} \citep{Savin2006}.
Our experimental work also motivated a series of state-of-the-art DR 
calculations for L-shell ions \citep{Badnell2003, Gu2003}.

Recently \cite{Chakravorty2008} have used the state-of-the-art theoretical
DR data of \cite{Badnell2003} and \cite{Badnell2006a,Badnell2006b} to
re-investigate this thermal instability.  They find that the new Fe L- and 
M-shell ion DR data results in ``a larger probability of having a 
thermally stable warm absorber at $10^5$~K than previous predictions and 
also a greater possiblity for multiphase gas.''

The reliability of the analyses reported above for the Fe M-shell UTAs
and for the thermal stability of warm absorbers at $\sim 10^5$~K depends on
the accuracy of the DR data used.  However, the two theorists
responsible for much of the available state-of-the-art theoretical
DR rate coefficients estimate that were they to carry out calculations
for M-shell iron ions in charge stages $6+$ and lower, then such data can
be uncertain by up to a factor of $10$ due to uncertainties in the
position of the relevant low-energy DR resonances
\citep{Badnell2008, Gu2008}.

In response to the clear need for reliable low-temperature Fe M-shell DR data, 
we are conducting a series of ion storage-ring merged beams measurements.
These are being carried out utilizing the \tsr heavy-ion
storage-ring \citep{Habs1989} located at the Max-Planck-Institute for 
Nuclear Physics in Heidelberg, Germany.
Using \tsr we have already carried out DR measurements for Mg-like 
\fefourteenplus \citep{Lukic2007}, Al-like \fethirteenplus \citep{Schmidt2006}, 
and Ar-like \feeightplus and K-like \fesevenplus \citep{Schmidt2008}.
Work predating the current project on Na-like \ion{Fe}{16} was carried out 
by \cite{Linkemann1995} and \cite{Mueller1999}.
A bibliographic compilation of storage-ring DR measurements with 
astrophysically relevant ions has recently been given by 
\cite{Schippers2008}.

DR is a two-step recombination process that begins when a free electron 
approaches an ion, collisionally excites a bound electron of the ion, and 
is simultaneously captured into a level with principle quantum number $n$.
The electron excitation can be labeled $Nl_j \rightarrow N'l'_{j'}$, where 
$N$ is the principal quantum number of the core electron, $l$ the orbital 
angular momentum, and $j$ its total angular momentum.
Conservation of energy requires for dielectronic capture that 
\begin{equation}
  \label{formel:energyconservation}
 \Delta E = E_k + E_b
\end{equation}
where $\Delta E$ is the core excitation energy, $E_k$ the kinetic energy 
of the incident electron,
and $E_b$ the binding energy of the captured electron in the core-excited 
ion.
Because $\Delta E$ and $E_b$ are quantized, DR is a resonant process.
We classify the core excitation by $\Delta N = N' - N$.
The intermediate state, formed by simultaneous excitation and capture, may 
autoionize. The DR process is complete when the intermediate state emits a 
photon which reduces the total energy of the recombined ion to below its 
ionization limit.
See \cite{Mueller2008} for additional details.

In this paper we report our recent results for measurements of $\Delta N=0$ 
DR in Cl-like \fenineplus and S-like \fetenplus and derive 
plasma rate coefficients for inclusion in astrophysical modelling codes.
The relevant channels for \fenineplus are
\begin{eqnarray}
\label{formel:FeXchannels}
{\rm Fe}^{9+}\ (3s^2\,3p^5\  [^2{P}^o_{3/2}]) + \eminus
  \rightarrow \left\{ \begin{array}{ll}
    {\rm Fe}^{8+}\ (3s^2\,3p^5\, [^2{P}^o_{1/2}]\, nl) \\
    {\rm Fe}^{8+}\ (3s\,3p^6\,nl) \\
    {\rm Fe}^{8+}\ (3s^2\,3p^4\,3d\,nl) \\
    {\rm Fe}^{8+}\ (3s\,3p^5\,3d\,nl)
    .
  \end{array} \right.
\end{eqnarray}
For the \fetenplus parent ion, the relevant capture channels are
\begin{equation}
\label{formel:FeXIchannels}
\mathrm{Fe}^{10+} (3s^2\,3p^4\ [^3{P}_{2}]) + \eminus
  \rightarrow \left\{ \begin{array}{ll}
    \mathrm{Fe}^{9+}\ (3s^2\, 3p^4\,[^3P_{1,0}; {^1}\hspace*{-0.1em}D_2; {^1}S_0]\, nl)\\
    \mathrm{Fe}^{9+}\ (3s\, 3p^5 \, nl)\\
    \mathrm{Fe}^{9+}\ (3s^2\, 3p^3\, 3d\, nl)\\
    \mathrm{Fe}^{9+}\ (3s\, 3p^4\, 3d\, nl)
    .
  \end{array}
  \right.
\end{equation}
In the energy range below approximately 90 (80)~eV the DR spectrum 
for \fenineplus (\fetenplus) is dominated by $\Delta N=0$ core excitations.
The energies of the core transitions relative to the ground 
levels of the respective parent ions are listed in Tables 
\ref{table:fe9energylist} and \ref{table:fe10energylist}.

The remainder of this paper is organized as follows:
Section~\ref{theory} gives a brief summary of the theoretical description.
Section~\ref{exp_setup} describes the experimental setup used in this 
work.
Section~\ref{results_mbrrc} shows our experimental results for the merged 
beams recombination rate coefficient (MBRRC) for S-like \fetenplus and 
Cl-like \fenineplus.
The experimentally derived DR plasma recombination rate coefficient (PRRC),
a comparison with theory, and
fitting results for each ion are also presented in Sec.~\ref{results_plasma}.
A summary is given in Sec.~\ref{conclusion}.

\section{Theory}
\label{theory}

\cite{Jacobs1977} published the first DR calculations for \fenineplus and \fetenplus 
using a ``no-coupling'' scheme, allowing only for dipole core 
excitations in the dielectronic capture process and paying no detailed attention 
to the low-energy resonances important in photoionized gas.
This was because their work was intended for high temperature plasmas  and 
computational limitations of the time necessitated truncating the 
calculations to keep them tractable.
\cite{ArnaudRaymond1992} incorporated the \cite{Jacobs1977} results and
adjusted them to the high-temperature behavior of the LS calculations
from \cite{Hahn1989}.

Stimulated by the AGN X-ray observations cited in Section~\ref{introduction} 
and by our experimental results for a number of M-shell iron ions,
\cite{Badnell2006b} revisited a series of Fe M-shell ions, including the
\fenineplus and \fetenplus ions.
Using the AUTOSTRUCTURE package \citep{Badnell1986} he performed multi-configuration Breit-Pauli (MCBP) calculations for Fe ions 
with valence $3p^{q}$ electrons $(q=0-6)$.
In this state-of-the-art theory, photorecombination, that is radiative 
recombination (RR) and DR, is calculated in the independent process 
isolated resonance approximation \citep{SeatonStorey1976}, i.e., 
RR is treated separately from DR.
The validity of this approach has been demonstrated by \cite{Pindzola1992}. 
Note that it should be implicitly understood thoughout that DR also 
comprises trielectronic recombination (TR) and possible higher order 
resonant processes wherever these are significant.
The importance of TR has been shown by \cite{Schnell2003a} in Be-like 
\ion{Cl}{14}, where TR contributes up to $40\%$ of the photorecombination 
plasma rate coefficient, depending on the plasma temperature.

Details of the MCBP DR calculations have been reported in \cite{Badnell2003}.  
Briefly, the AUTOSTRUCTURE code was used to calculate energy levels as 
well as radiative and autoionization rates in the intermediate coupling 
approximation. These must be post-processed to obtain the final-state 
level-resolved and total DR data. The energy levels were shifted to 
known spectroscopic values for the $3 \rightarrow 3$ transitions.  
Radiative transitions between autoionizing states were accounted for in 
the calculation. The DR cross section was approximated by the sum of 
Lorentzian profiles for all included resonances.
The basic configurations which describe the $N$-electron target for 
\fenineplus are $3s^2\,3p^5$, 
$3s\,3p^6$, $3s^2\,3p^4\,3d$, $3s\,3p^5\,3d$, $3s^2\,3p^3\,3d^2$, and 
$3s^0\,3p^6\,3d$.
The \fetenplus target configurations are $3s^2\,3p^4$, $3s\,3p^5$, 
$3s^2\,3p^3\,3d$, $3p^6$, $3s\,3p^4\,3d$, $3s^2\,3p^2\,3d^2$, and 
$3s^0\,3p^5\,3d$.
A closed-shell Ne-like core is assumed for each ion.
Both configuration sets are identified as ``7CF'' in 
\cite{Badnell2006b}.

For \fenineplus we derived a theoretical merged beams DR rate coefficient 
by reproducing the earlier calculations of \cite{Badnell2006b}.  For 
\fetenplus
we have carried out new DR calculations using the state-of-the-art MCBP 
theoretical method as described in \cite{Badnell2006b}.
In both calculations only the strong $\Delta N=0$ DR channels associated 
with $3\rightarrow 3$ transitions have been considered. The weaker $\Delta 
N \ge 1$ DR channels have been neglected in order to keep the calculation 
managable. For the $3p^q$ systems considered by \cite{Badnell2006b}, 
their contributions to the DR PRRC have been estimated to be 
insignificant for photoionized plasmas and less than $10\%$ of the $\Delta 
N = 0$ rate coefficient for collisionally ionized plasmas.

Revisiting the \fetenplus calculations performed in \cite{Badnell2006b} 
turned out to be necessary as initially the theoretical energy structure 
of the Rydberg series limits at around $65$~eV did not coincide with the
experimental results.
\cite{Badnell2006b} calculated the levels of core excitations for all 
configurations in the ``7CF'' list. Those results were shifted to match 
the tabulated level-data of \cite{AtomicSpectraDatabase}. However, 
for the \fetenplus $3s^2\,3p^3\,3d$ configuration the 
tabulated levels are incomplete and \cite{Badnell2006b} shifted all of the
roughly 50 levels of this configuration by the average shift derived from the 
12 tabulated levels.
Here we shift these 12 known levels 
to their explicit positions given by \cite{AtomicSpectraDatabase} and use the 
average shift only for the remaining levels.
This improves the agreement between theory and experiment for the energy
structure at the $3s^2\,3p^3\,3d$ series limits.
The resulting new theoretical plasma rate 
coefficient changes by less then $3\%$ from the \cite{Badnell2006b} data 
in the regimes where \fetenplus is abundant in photoionized or 
collisionally ionized plasmas.

RR calculations for iron M-shell ions were also carried out by 
\cite{Badnell2006b} using AUTOSTRUCTURE.
Here, we use his RR cross section results for both \fenineplus and 
\fetenplus.

\section{Experimental Setup}
\label{exp_setup}

Measurements were performed utilizing the heavy-ion storage-ring \tsr.
The experimental method used has been described in detail by 
\cite{Kilgus1992}, \cite{Lampert1996}, \cite{Pastuszka1996}, 
\cite{Schippers2001}, \cite{Wolf2006}, \cite{Lestinsky2008}, and 
\cite{Schmidt2008}.
The key components are discussed briefly here.
See, in particular, \cite{Schmidt2008} for details not touched on here.

The \tsr facility is uniquely configured with two separate electron-ion 
merged beams sections.
This instrumentation allows us to perform continuous electron cooling 
\citep{Poth1990} in one straight section and to carry out measurements of the 
desired electron-ion collisions in another section.
The first electron beam apparatus merges a magnetically guided cold 
electron beam with the ions for a distance of $\approx 1.5$~m. This 
co-propagating electron beam serves as a heat bath for the ions 
\citep[hereafter the electron cooler;][]{Steck1990}.
The second apparatus operates as a dedicated collisional probe for
measurements at tunable relative velocity \citep[hereafter the electron 
target or probe beam;][]{Sprenger2004}.
It employs schemes for adiabatic transverse expansion 
and adiabatic acceleration to achieve low transverse and longitudinal 
temperature spreads of the probe beam. 

The physical separation of the cooling and probe beam functions also has the 
benefits of improving the ion beam quality by allowing for constant 
cooling of the ions.
Simultaneous cooling while probing stabilizes the ion beam energy.
The drag force \citep{Wolf2006} exerted on the ion beam by the probing 
electron beam are offset by the cooling forces action in the ions in the 
cooler.
This improves the achievable resolution as well as the precision 
of the energy scale at low relative energies.

\label{detector}
In the recombination process the momentum transfer is small and the 
recombined ions insignificantly alter their initial trajectory.
However, due to their decreased charge state, they are deflected by the 
first dipole magnet downstream of the electron target onto a trajectory 
different from the parent ions and directed onto 
a YAP:Ce crystal \citep{Baryshevsky1991} used as a fast scintillator 
which can be positioned within the vacuum system of \tsr to collect a 
desired product ion species \citep{Wissler2002, Lestinsky2007}.
The impacting ions deposit their complete kinetic energy in the crystal 
and stimulate the emission of hundreds of photons within a few 
nanoseconds.
These light pulses are coupled through a window onto  a photomultiplier 
tube, and its output signal is amplified and discriminated.
The discriminator output pulses are counted in a scaler channel of the 
data acquisition computer controlling the experiment.
By stepping the detector position horizontally and vertically through the 
recombined ion beam the width of the product beam was determined to 
$\approx 1$~mm in each direction, which basically is an image of the 
stored cooled
parent beam. With a detector size of $20~\mathrm{mm}\times20~\mathrm{mm}$, 
all recombined ions created in the electron-ion overlap region are collected by 
the active detector surface after aligning the center of the detector to 
the center of the product beam.
The detector gives a signal which consists almost entirely of
recombined target ions; cosmic ray incidents are observed with a rate of
$<1$~Hz.
With a pulse height spread of $\approx 5\%$, the discriminator cut-off can 
be set safely without suppressing real events.
The total count rate was less than $100$~kHz at all times. The dead time 
of the detector was determined by the discriminator pulselength of 50~ns. 
Thus, we assume an upper limit for dead-time induced loss of events to 
below $0.5\%$.
Hence we can exclude electronic or geometric cut-offs to the detection 
efficiency and assume a detector efficency of unity.

Along the trajectory from the electron-ion interaction region to the 
detector, the recombined ions experience various magnetic fields, 
e.g., due to the demerging section of the electron target, focussing elements 
in the \tsr beamline, and the first dipole magnet downstream of the 
target.  These magnetic fields induce motional electric fields as seen by 
the ions, leading to field ionization of high-$n$ Rydberg levels.
For the magnetic field strengths and the ion velocities used, the 
semi-classical value of this cut-off \citep{Gallagher1994}
is $n_{\mathrm{cut}} = 39$ for \fenineplus and $41$ for \fetenplus.
However some of the $n$ levels above $n_{\mathrm{cut}}$ can radiatively 
decay to below $n_{\mathrm{cut}}$ during the flight time from the target 
through the magnets and are thus detected.
These ionization and radiative de-excitation processes have been described 
in detail by \cite{Schippers2001}, and are quantitatively well understood 
with $(n,l)$-specific modelling of the de-excitation probabilities.
We use this approach when comparing our MCBP results to our 
experimental results.

Data collection was performed by repeating a measurement cycle consisting 
of successive phases of injection, electron cooling, and energy scans.
For the present work, \fenineplusc and \fetenplusc beams of $93.2$  and 
$115$~MeV, respectively, were prepared in a tandem accelerator, charge- 
and mass-selected (atomic weight $A=56$), and injected into \tsr using 
multi-turn injection and ecool stacking \citep{Grieser1991}.
During the injection phase, ions were fed into \tsr in pulses once per 
second.
Simultaneous electron cooling decreased the spatial and velocity spreads 
of newly injected ions, freeing up storage-ring acceptance phase space 
volume for the following injection shot.
Accumulation of ions over thirteen and ten injections were required per 
cycle for \fenineplusc and \fetenplusc, respectively, to establish 
time-averaged mean ion currents of $\approx 25~\mu\mathrm{A}$ in the 
storage-ring during the
active data collection phase.
After the injection phase, continuous electron cooling was applied for 
$4.5$~s to bring the average laboratory ion velocity
$\left< v_{\mathrm{ion}} \right>$ to the average laboratory electron 
velocity $\left< v_e \right>$, thereby decreasing the ion beam momentum 
spread to equilibrium at $\Delta p/\left<p\right> \approx 10^{-4}$.

The stripping stages of an accelerator, using beam foil stripping 
targets, are known to produce ions in metastable states 
\citep{MartinsonGaupp1974}.
For \fenineplus, the only metastable level is the $3s^2\,3p^5\, ^2P_{1/2}$, 
which decays to the $^2P_{3/2}$  ground level with a lifetime of $\approx 
15$~ms \citep{AtomicSpectraDatabase}.
Within the initial $4.5$~s cooling period after injection and before the 
energy scans were started, the population of the \fenineplus metastable level 
decayed to an insignificant fraction.
\fetenplus in its $3s^2\,3p^4$ ground configuration has $^3P_1$, $^3P_0$,  
$^1D_2$, and $^1S_0$ metastable levels lying above the $^{3}P_2$ ground 
level.
The $^3P_0$ level has a lifetime of about $4$~s; the lifetimes of the 
other metastable levels are at least an order of magnitude shorter
\citep{AtomicSpectraDatabase}.
Assuming that the ion beam is purely in the $3s^2\,3p^4$ configuration and each 
level is populated with a Boltzmann distribution, we have modelled the 
decay chain network to derive the evolution of the relative populations
during the course of the measurement.
Using this model and taking into account the ecool-stacking injection 
pattern, the calculated metastable fraction is below $2\%$ at the 
beginning of the measurement energy scan.
Scaling the MCBP calculations for the $^3P_0$ metastable level down by
the population ratio indicates that there are no significant resonances
expected in our measured MBRRC spectrum.  The MCBP PRRC calculations for
the metastable level are $\approx 15\%$ smaller than the PRRC ground
state results in the photoionized regime and $\approx 35\%$ smaller in
the collisionally ionized regime.  Scaling these down by the population
ratio gives a smaller than 1\% error in our experimentally-derived PRRC
for the ground state.  Thus the largest uncertainty from the metastable
fraction is the uncertainty in the ground state population of the ion
beam due to the possible 2\% contamination of beam.  We treat this as a
systematic uncertainty which we add in quadrature to the total
experimental error.

The total energy range studied was subdivided into multiple energy 
intervals and measured in separate runs.
The energy range of each interval was selected to overlap in the 
laboratory frame by $50\%$ with adjacent energy scans.
In total, the electron lab energy was varied
for \fenineplus from $857.7$ to $1584.1$~eV and
for \fetenplus from $1098.9$ to $1823.3$~eV,
with the electrons matching the velocity of the ions at an energy $E(v_{\rm rel} = 0) \approx 915$~eV for 
\fenineplus and $\approx 1132$~eV for \fetenplus.
The lab energy intervals were subdivided into 800 steps with a
stepsize of $0.122$~eV.
These were scanned in a wobble mode, where each measurement step of the 
scanning electron beam energy was followed by a  reference step at a fixed 
energy.
Slew-rates in associated power supplies were taken care of by an 
additional 5~ms waiting time before beginning the data collection after 
each energy jump.
Data were then collected at each step for 10~ms.
Within 24~s the energy scan was completed and a new cycle started over 
with a new injection phase.
A run required one hour on average to accumulate sufficient counting 
statistics.
A total of 19 (14) individual runs for \fenineplus (\fetenplus) were 
collected, covering the relative collision energy range from $0$ to about 
92 (83)~eV.

The reference step serves as a measurement of the background rate 
primarlily due to electron capture from the residual gas.
Reference points have been chosen in flat, unstructured regions of the 
cross section near the energy scan interval, to suppress artifacts 
introduced by large energy jumps.
Despite these precautions the background level still varied between 
overlapping runs.
We derived and subtracted these offsets by successively comparing the 
integrated MBRRC signal in the overlapping regions of neighboring runs.
The associated statistical uncertainties in the integrated signal 
contribute to the
error budget.
We began with the runs spanning across the highest observed DR series 
limits, using center-of-mass reference energies at $E_{\mathrm{ref}} = 
73.19$~eV for \fenineplus and $70.64$~eV for \fetenplus, as no DR signal 
was seen or is expected at these energies.
The background above the series limits for these runs was shifted by 
adding the appropriate amount to match
the small RR signal, as given by the calculations of \cite{Badnell2006b}
and modified to take the experimental field ionization into account.

The relative collision energy is established by a transformation to the 
center-of-mass frame described by \cite{Kilgus1992}. 
Precise knowledge of the laboratory electron energy $E_e$ and ion energy $E_i$ 
is necessary.
The electron energy is measured with a highly sensitive HV probe and 
corrected for space charge effects in the electron beam, analoguous 
to \cite{Kilgus1992}.
The ion energy is defined by the electron cooling condition 
$\left<v_i\right> = \left<v^{\mathrm{cool}}_e\right>$. 
The electron energy at cooling, $E^{\mathrm{cool}}_e$ is inferred from 
the constraint that the center-of-mass MBRRC spectrum scanned for $v_e < v_i$ and 
$v_e > v_i$ must be symmetric around $v_e = v_i$.
From such measurements we determined 
$E^{\mathrm{cool}}_e = 914.70(3)$~eV for \fenineplus and 
$1131.57(3)$~eV for \fetenplus.
From here on and throughout this paper a quantity in parentheses specifies the 
uncertainty on the last significant digit(s). All errors are 
given at a confidence interval of $90\%$, unless otherwise noted.
\label{confidenceinterval}

The resulting center-of-mass energy scale was linearly adjusted by a
small correction 
factor to match the theoretical calculations. To derive the 
calibration factors we selected distinct high-$n$ Rydberg resonances with 
narrow structure and sufficiently clean surroundings and scaled the 
experimental energy axis to match the theoretical MBRRC energy structure. 
For \fenineplus, we used the peak at $57.4$~eV, attributed to 
the $3s^2\,3p^4\,(^3P)\,3d\,10l$ resonance in the \feeightplus product spectrum. 
For \fetenplus we used the peak at 
$49.05$~eV, attributed to the $3s^2\,3p^3\,(^2D^{\mathrm{o}})\,3d\,9l$ 
resonance in the \fenineplus product spectrum.
Based on these we calibrated the energy axis for \fenineplus 
by scaling it up with a factor of $1.0006(6)$ and for \fetenplus by $1.0065(7)$.

The experimental energy distribution is best described as a flattened 
Maxwellian distribution \citep{Kilgus1992} which is characterized by the longitudinal and 
transverse temperatures $\ktpar$ and $\ktperp$.
The experimental resolution $\Delta \hat{E}$ at collision energy $\hat{E}$ 
is approximately given by $\Delta \hat{E} = [(\ln(2)\ktperp)^2 + 
16\ln(2)\hat{E}\ktpar]^{1/2}$
\citep{Mueller1999}.
To reduce $\ktperp$, the target electron beam was adiabatically expanded 
\citep{Pastuszka1996} by a factor of 30.
The transverse temperature derived from fitting the measured DR spectrum 
was $\ktperp^\mathrm{fit} = 3.6(3)$~meV for both ion species (see
Sec.~\ref{prrc_derivation} for details on the fitting).
This is consistent with expectations of $\ktperp = 3.6(1)$~meV for thermal
emission from a cathode with temperature $\kt_\mathrm{cath} \approx 110(5)$~meV 
and an adiabatic transverse expansion factor $\xi_e = 30$.
The longitudinal temperature was derived from the width of distinct 
high-$n$ resonance features in the Rydberg series which yields $\ktpar = 
38(1)~\mu{\mathrm{eV}}$.
With these temperatures the experimental energy spread for both ions 
amounts to
$\Delta \hat{E} \approx 0.003$~eV at a collision energy of $\hat{E}=0.001$~eV,
$0.005$~eV at $0.01$~eV,
$0.009$~eV at $0.1$~eV,
$0.023$~eV at $1$~eV,
$0.067$~eV at $10$~eV, and $0.21$~eV at $100$~eV.

The dominant systematic uncertainties in the experimental results arise 
from the ion current measurement and the length of the interaction region 
and are estimated to be $25\%$ \citep{Lampert1996}.
The alignment of the electron beam angle and perpendicular position with 
regard to the ion beam axis has an uncertainty of below 1~mm throughout 
the $\approx 1.5$~m long interaction zone.
With an expansion factor of $\xi_e = 30$, the electron beam has  a 
diameter of 9~mm. Thus, the $\approx 1$~mm wide ion beam is fully
embedded within the electron beam in this region.
An uncertainty of $3\%$ arises from variations in the transverse density 
profile of the probe electron beam.
For \fetenplus we add $2\%$ to account for the possible $^3P_0$ metastable 
fraction.
Adding all contributions in quadrature yields an estimated total 
uncertainty of $25\%$ for both ions for the MBRRC.
Another contribution to the error budget results from  corrections to the
nonresonant background due to the shifting of the background level between
scans.
This uncertainty in the shift amounts to less than $2.7\times 
10^{-11}$~cm$^3$s$^{-1}$ for \fenineplus (corresponding to an average of 
$8\%$ of the integrated MBRRC in the energy range from 2 to 60 eV).
For \fetenplus, it is below $1.2\times 10^{-11}$~cm$^3$s$^{-1}$ or 3\% in 
this range.
The contributions are insignificant outside this interval.
For the derived Maxwellian PRRC this background correction when integrated 
with a Maxwellian distribution is insignificant at low temperatures and 
less than $8\%$ ($3\%$) for \fenineplus (\fetenplus) at temperatures above 
$10^5$~K.

\section{Experimental and Theoretical Results}
\label{results_mbrrc}

\begin{figure}
    \plotone{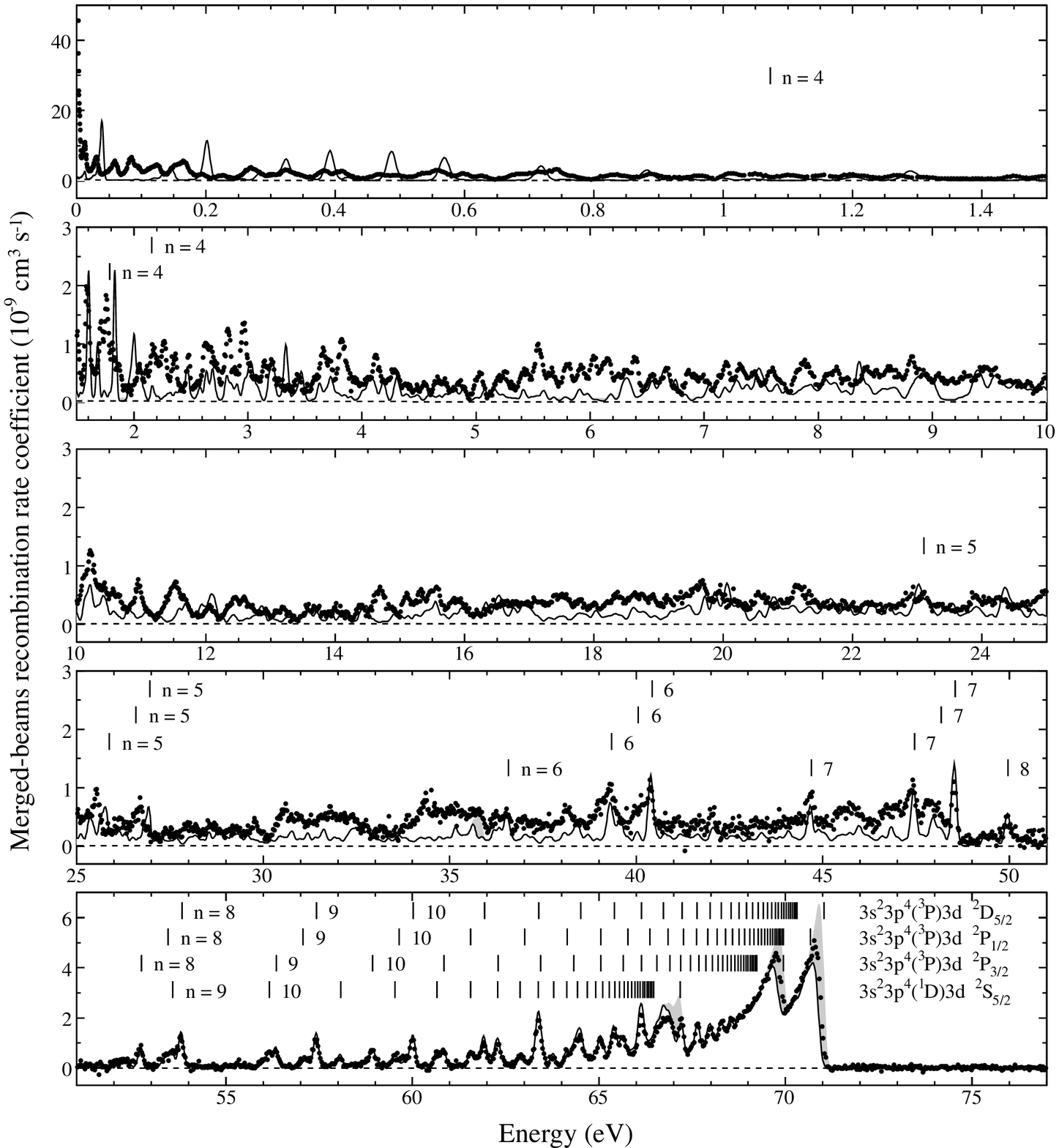}
    \caption{\label{fig:fe9_mbrrc}\fenineplus MBRRC as a function of
	relative energy.
	The experimental MBRRC data is shown with filled circles.
	Our AUTOSTRUCTURE calculations including field ionization are 
	shown by the solid line. The additional theoretical high-$n$ contributions
	missing from the experimental data due to field ionization are
	shown by the grey shaded areas.
	The theoretical rate coefficient has been scaled down by a uniform 
	factor of $1.46$ (see Sec.~\ref{fe9mbrrc}).
	Vertical lines indicate the most clearly visible Rydberg series and are
	labeled by the corresponding core excitation. See text for 
	details.
    }
\end{figure}

\begin{figure}
    \plotone{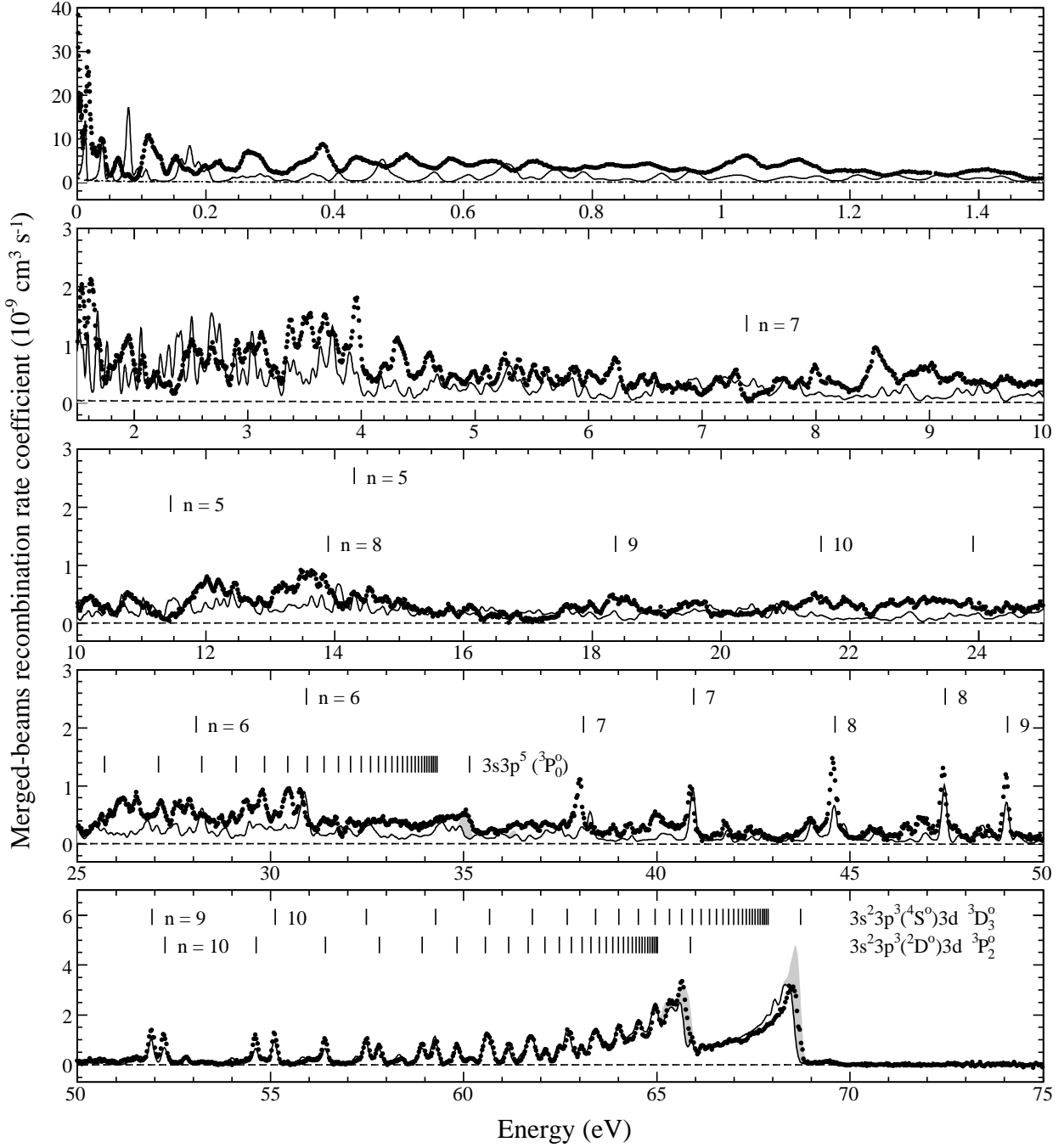}
    \caption{\label{fig:fe10_mbrrc}
	Same as Fig.~\ref{fig:fe9_mbrrc} but for \fetenplus{}.
	Here, the theoretical rate coefficient has been scaled down by a uniform
	factor of $2.00$ (see  Sec.~\ref{fe10mbrrc}).
    }
\end{figure}

Figs.~\ref{fig:fe9_mbrrc} and \ref{fig:fe10_mbrrc} show the experimentally
obtained MBRRC for \fenineplus and \fetenplus as a function of collision 
energy (circles).
Also shown are the results of our DR+RR MCBP calculations using AUTOSTRUCTURE.
These cross section calculations have been transformed into an MBRRC
by convolving them with the relative velocity and the flattened,
double-Maxwellian electron velocity distribution of 
the experiment using the thermal spreads derived from the measured data.
The solid line shows the calculated MBRRC results taking into account 
the $(n,l)$-specific modifications due to field ionization, as discussed 
in more detail in Sec.~\ref{prrc_derivation}.
The grey areas indicate the contributions for capture into Rydberg states
up to $n=1000$ which are missing from storage-ring results due to
field ionization.
The dashed lines show the RR only contributions \citep{Badnell2006b}, 
convolved with the experimental resolution and modified by the field 
ionization model. 

Vertical lines
are used in Figs.~\ref{fig:fe9_mbrrc} and  \ref{fig:fe10_mbrrc} to indicate
the energy position of resonances from the dominant Rydberg series.
These resonance energy sequences have been calculated up to $n_\mathrm{cut}$
using equation~\ref{formel:energyconservation} and a hydrogenic 
approximation for the binding energy $E_b$.
A final line is drawn at the energy position of the $n = \infty$ limit.
The series limit $\Delta E$ for each of the respective series
for \fenineplus and \fetenplus were taken from
tables \ref{table:fe9energylist} and \ref{table:fe10energylist}, respectively,
and are labeled by the corresponding core excitation level.

\subsection{\fenineplus merged-beams recombination rate coefficients}
\label{fe9mbrrc}

The \fenineplus spectrum is rich in DR features.
At energies from $0$ to $49.7$~eV we observe broad, unresolved structure, 
which can be ascribed to the superposition of many DR resonances, but in 
this energy region
it is impossible to reliably attribute the observed features to 
specific resonances.
The agreement of our MCBP calculations with our measured DR structure is 
marginal.  The calculated resonances barely coincide with the measured 
features. Still, suggestions of Rydberg series limits can be seen in the 
experimental data at 
some of the expected energies for such features, notably $35.9$ and 
$48.7$~eV, representing the $3s\,3p^6\ ^2S_{1/2}$ and
$3s^2\,3p^4\,(^3P)\,3d\ ^4D_{\{7/2,\,\dots\,1/2\}}$ core excitations, respectively
(see Tab.~\ref{table:fe9energylist}).

Above $48.7$~eV  the measured spectrum becomes less crowded and DR 
resonances can be more reliably assigned to observed peaks.
Resonances due to higher excitations are particularly clear for electron
capture into levels starting at $n=6$ and can be followed up to 
approximately $n=13$
before they blend into the characteristic cusp approaching the series 
limit.
These resonances are associated with the
$3s^{2}\, 3p^{4}\, (^{1}\hspace*{-0.1em}D)\, 3d\ ^{2}\hspace*{-0.1em}S_{5/2}$,
$3s^{2}\, 3p^{4}\, (^{3}\hspace*{-0.1em}P)\, 3d\ ^{2}\hspace*{-0.1em}P_{\{3/2,\,1/2\}}$, and
$3s^{2}\, 3p^{4}\, (^{3}\hspace*{-0.1em}P)\, 3d\ ^{2}\hspace*{-0.1em}D_{5/2}$ core excitations.
The limits arising from these series are found in the experimental data at 
approximately $67$~eV, $70$~eV, and $71$~eV, respectively.

It is useful to compare in detail the integrated total DR resonance
strengths over various energy intervals.
The measured data at very low colllision energies below $E \lesssim \ktperp$ (here $\approx 3$~meV)
contain uncertain contributions due to RR rate enhancement further 
discussed in Sec.~\ref{prrc_derivation}.
In the energy interval from $0.003$ to $0.5$~eV
we find a ratio of the theoretical to the experimental strengths 
of $\kappa = \int\alpha_{\mathrm{theo}}dE/\int \alpha_{\mathrm{exp}}dE = 
0.97(12)$.  At energies from
$0.5$ to $2$,
$2$ to $10$,
$10$ to $36$, and
$36$ to $48.7$~eV,
we observe a consistently stronger experimental MBRRC than predicted by 
theory,
with similar ratios of $\kappa = 0.67(2)$, $0.69$, $0.76$,
and $0.70$, respectively.
These ratios, similar among each other for these energy ranges, carry 
errors significantly below the least significant digit given, unless 
explicitly stated.
From $48.7$~eV and beyond, where the resonance structure of the 
calculations agrees well with the experiment, we find that 
the measured MBRRC is nevertheless considerably weaker than predicted by MCBP
theory.
The theory-to-experiment ratio amounts to $\kappa=1.46$ in the range of 
$48.7$ to $66$~eV and to $\kappa = 1.38$ at $66$ to $72$~eV; as the latter 
range spans across three pronounced series limits, the theoretical data 
there are also sensitive to the uncertainties contained in the modeling of 
the field ionization effect. Integrated resonance strengths for the 
various energy ranges are given in Table~\ref{table:fe9integrals}.
In the comparison of \figurename~\ref{fig:fe9_mbrrc} and for estimating 
the high-Rydberg contributions unobserved because of field ionization, we 
use theory scaled down by $\kappa$ from the range of $48.7$ to $66$~eV, 
yielding good match for the lower Rydberg resonances.

\subsection{\fetenplus merged beams recombination rate coefficients}
\label{fe10mbrrc}

The \fetenplus DR spectrum is as rich as that of \fenineplus.
Below about $35$~eV, no interval can be found where the experimental spectrum
is free of DR contributions and broad, unresolved structure dominates the 
spectrum.
Again, we find in this range poor agreement between our measurement and 
our theoretical calculation of the resonance structure. 
Near $35.1$~eV a weak series limit can be identified, showing a broad, 
unstructured cusp in the DR signal.
This is attributed to DR via $3s\,3p^4\,(^3P^o_0)$ core excitation (see 
Tab.~\ref{table:fe10energylist}).
Yet, we are unable to unambiguously identify any individual resonance features
due to this series.

Above 35~eV the measured spectrum becomes less crowded and DR resonances 
can be assigned to observed peaks with more confidence.
DR resonances via $3s^2\,3p^3\,(^2D^o)\,3d\,^3P^o_2$ and
$3s^2\,3p^3\,(^4S^o)\,3d\,^3D^o_3$ core excitations
can be clearly identified, with series limits at $65.9$~eV and $68.7$~eV, 
respectively.
Distinct resonance features due to these two series can be traced for 
capture
into Rydberg levels from $n = 7$ to $n \approx 17$ before the resolving 
power of the experiment blends them into one another.
For the $n=5$ and $6$ levels of these series, unambiguous identification
of the corresponding peaks in the measured spectrum
is not possible due to a combination of fine structure splitting and
resonances from other series.
At $69.7$~eV another small cusp in the data  can be found which can be 
attributed to $3s^2\,3p^3\,(^4S^o)\,3d\,^3D^o_2$ core excitation.

Comparing in detail the integrated theoretical and experimental resonance 
strengths for DR over various energy intervals we find a 
behavior different from what was found for \fenineplus, with the 
integrated resonance strengths and the theory-to-experiment ratios 
$\kappa$ listed in Table~\ref{table:fe10integrals}. Also here, theory 
generally underestimates the experimental MBRRC by up to about $30\%$, 
while it overestimates it by up to $100\%$ along the upper part of the 
dominant Rydberg series.
In the comparison of \figurename~\ref{fig:fe10_mbrrc} and for estimating
the high-Rydberg contributions unobserved because of field ionization,
we use theory scaled down by $\kappa = 2$ from the range of $55.5$ to $65$~eV.

\section{Plasma DR rate coefficients}
\label{results_plasma}

\subsection{Derivation}
\label{prrc_derivation}

The derivation of a  plasma rate coefficient from the experimental MBRRC data
requires accounting for three issues relating to the measurement 
technique. These are broadening of 
resonances due the experimental resolution \citep{Schippers2004}, 
low-energy enhancement of the RR signal \citep{Gwinner2000, Hoerndl2006},
and field ionization of high-$n$ Rydberg levels.
In order to address these issues we split the measured MBRRC spectrum into 
two parts, as has been described more comprehensively
in the work of \cite{Schmidt2008}: one below $0.1$~eV and one above.

\begin{figure}
    \plotone{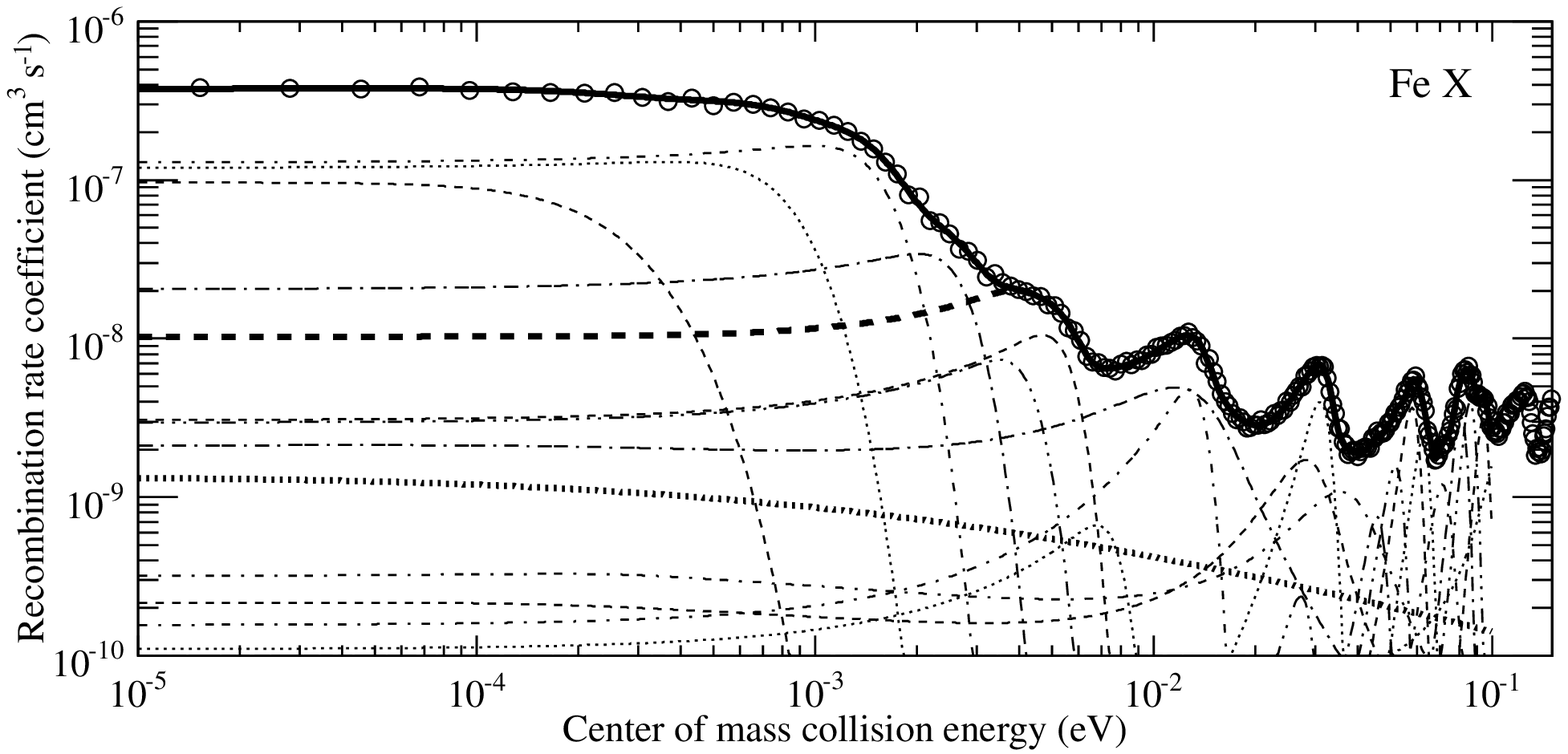}
    \plotone{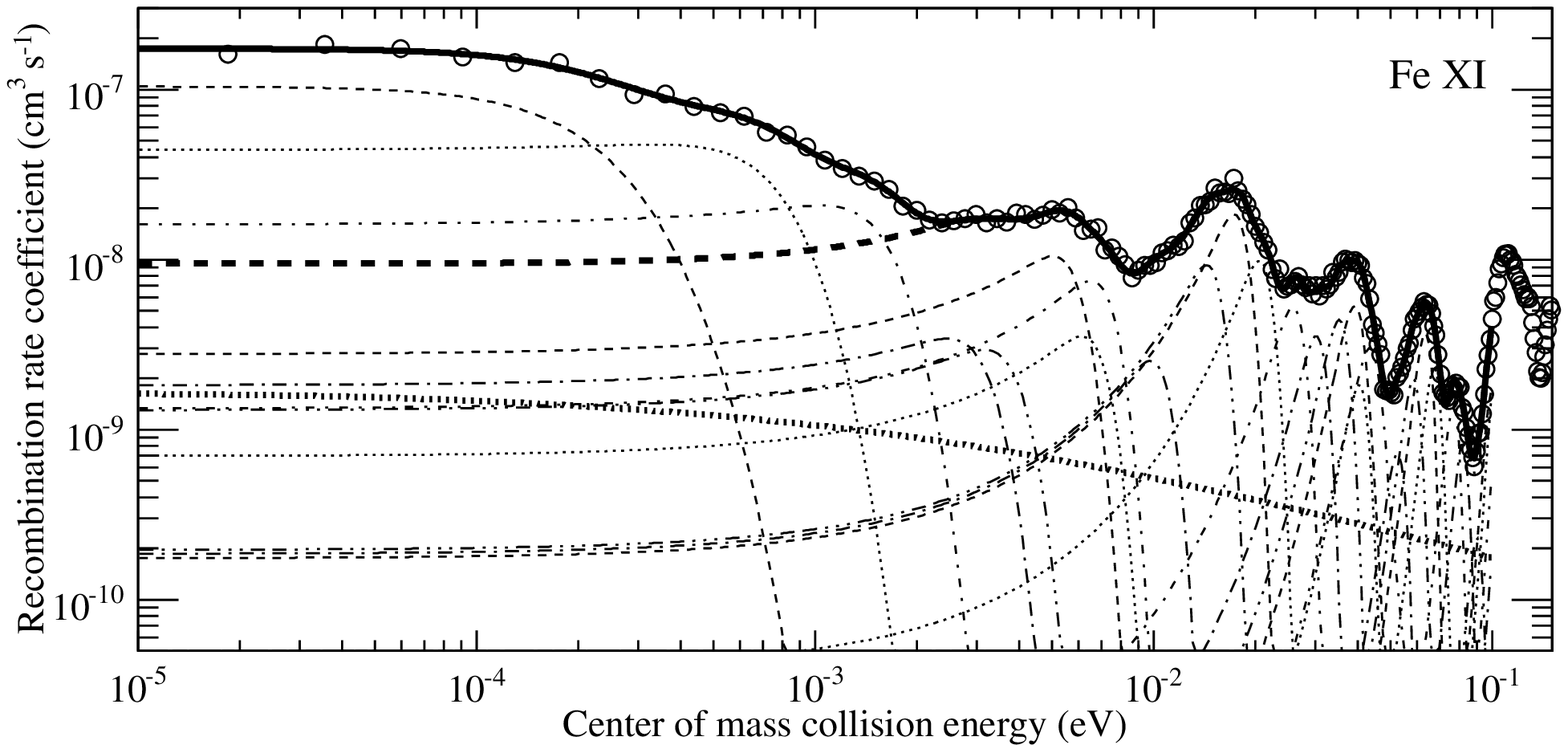}
    \caption{\label{fig:fits}Fitted MBRRC model spectrum in the energy 
	range from  $10^{-5}$ to $0.1$~eV for both \fenineplus (upper panel)
	and \fetenplus (lower panel).
	The open circles show the measured data. The thick solid line 
	shows the convolved fitted model spectrum
	$\left<v \sigma^{\rm DR}_{\rm lo}(E)\right>$ + $\left<v \sigma^{\rm RR}(E)\right>$.
    	The contributions from the individual resonances in the total model 
    	spectrum are indicated using thin lines and varying line style.
    	The thick dashed line gives the modified fit, where resonances below 
    	$3$~meV have been excluded.
	RR is shown with a thick dotted line.
    }
\end{figure}

For the measured data in the interval below $0.1$~eV, the DR PRRC is 
derived from an empirical model cross section spectrum $\sigma^{\rm DR}_{\rm 
lo}(E)$ fitted to the measured MBRRC.
This model spectrum is convolved with the relative velocity and a
flattened, double-Maxwellian electron velocity distribution 
\citep{Kilgus1992}.
The transverse and longitudinal temperatures, as well as
the resonance energies and strengths are free parameters in this fit.
The measured spectrum is a blend of an unknown number
of resonances which we model using a mixture of $\delta$-functions
and Lorentzian lineshapes.
For \fenineplus our model includes 21 resonances between 0 and $0.1$~eV. 
For \fetenplus the model uses 24 resonances in this energy range.
A few additional resonances just above the high end of the fitting interval 
are included to ensure the continuity of the fit with the measured data.
The curves for the total fit and for each resonance are shown in 
\figurename~\ref{fig:fits}.
Calculation of the RR cross section $\sigma_\mathrm{RR}(E)$ is modelled using
the hydrogenic Bethe-Salpeter description \citep{Hoffknecht2001}.
We scale the convolved RR rate coefficient $\alpha_\mathrm{RR}(E)$
to the results of the more sophisticated AUTOSTRUCTURE calculations
by \cite{Badnell2006b}.
The deviation of the scaled Bethe-Salpeter curve from AUTOSTRUCTURE
results is less than $1.5\%$ in the fitted energy interval.
Since these non-enhanced RR contributions (see below) account for only
about $1\%$ of the total MBRRC 
signal at 0~eV relative energy, this deviation can be neglected in the 
error budget.

In merged electron-ion beam experiments using magnetically guided electron 
beams one typically finds an enhanced MBRRC at  relative energy of the 
beams in a range starting below the transverse electron energy $\ktperp$, 
with $\ktperp \approx 3$~meV in the present experiment.
This is caused by the motionally induced electric fields generated
by the toroidal magnetic fields in the merging and demerging
regions of the electron coolers. 
\cite{Gwinner2000} and \cite{Hoerndl2006} have shown dependencies
on the electron beam temperature, ion species, and the specific geometry
of the electron beam setup.
\cite{Hoerndl2006} showed that the enhancement in general becomes visible
at energies $E \lesssim \ktperp$.
For measurements using the \tsr electron cooler, typical enhancement 
ratios in the range of $[\Delta \alpha(E)/\alpha_0(E)] + 1 = 1.5-3$
\citep{Wolf2003} 
were observed, where $\alpha_0(E)$ is the theoretically expected value
and $\Delta \alpha(E)$ is the contribution due to enhancement
\citep[for details see][]{Gwinner2000}.
This uncertainty in the low-energy recombination signal at interaction 
energies in the range below $\ktperp$ is inherent to our method,
however it does not contribute to the recombination signal in ``natural''
astrophysical plasmas and hence needs to be removed before
deriving a true plasma recombination rate coefficient.

Our modelling does not account for the RR enhancement and is thus likely 
to overestimate the strengths of the DR resonances at $E \lesssim 
\ktperp$, corresponding to a range below $3$~meV in the present 
experiment.
For assessing the possible systematic errors due to this uncertainty at 
the lowest merged-beam collision energies, 
we consider our results for two cases to derive an estimated upper 
limit and a hard lower limit for the DR contributions to 
$\alpha_\mathrm{exp}(E)$:
In the first case, we keep all 21 (24) fitted
resonances for \fenineplus (\fetenplus),
in the second case we remove the 4 (3) resonances below $3$~meV for
\fenineplus
(\fetenplus) from the model cross section spectrum.
Comparing the convolved model for the second case (thick dashed line in 
\figurename~\ref{fig:fits}) to the measured data, at $0.01$~meV we find
ratios of $\alpha_{\rm exp}(E)/\alpha_{\rm fit}^{wo}(E) \approx 36.6$ for 
\fenineplus and $\approx 18.3$ for \fetenplus as upper limits for the  
rate enhancement.
These enhancements are more than an order of magnitude larger than those 
determined for the cooler.
For the electron target a preliminary study of the rate enhancement found 
an enhancement ratio of $\approx 4.6$ for Li-like \ion{F}{7}
\citep{Lestinsky2007}.
\cite{Schmidt2008} also carried out measurements using the electron 
target. They reported enhancement factors of $7.5$ and $3.7$ for 
\fesevenplus and \feeightplus, respectively.
Thus, the amount of observed enhancement here strongly suggest the presence 
of unresolved DR resonances below $3$~meV.

We derive DR plasma rate coefficients
$\alpha_\mathrm{P,lo}^\mathrm{wi}(T)$ and 
$\alpha_\mathrm{P,lo}^\mathrm{wo}(T)$
by convolving the respective model DR cross sections 
(with and without those resonances below $3$~meV, respectively)
with the relative velocity and an isotropic Maxwellian electron
energy distribution \citep{Schippers2001}.
The average of $\alpha_\mathrm{P,lo}^\mathrm{wi}(T)$ and 
$\alpha_\mathrm{P,lo}^\mathrm{wo}(T)$ is taken as the experimentally
derived plasma DR rate
coefficients $\alpha_\mathrm{P,lo}(T)$. 
The half difference between both
is propagated
to the error budget as an experimental uncertainty associated
with the rate enhancement.
This error is large only at very low plasma temperatures and is unimportant 
for temperatures above $10^3$~K.

Above $0.1$~eV the measured data are treated by subtracting
the theoretical (small) RR from the experimental MBRRC spectrum  
to obtain a pure DR spectrum $\alpha_{\rm DR}(E)$.
The missing high-$n$ Rydberg contributions are accounted for 
by replacing the experimental data $\alpha_{\rm DR}(E)$
in intervals sensitive to field ionization with the scaled theoretical MBRRC
for $n\le1000$,
\begin{equation}
\alpha'_{\rm DR}(E) = \left\{
		\begin{array}{ll}
			\alpha_{\rm DR}(E) & : 0.1~{\rm eV} \le E \le E_{\rm cut},	\\
			\kappa^{-1}\alpha^{\rm MCBP}_{\rm DR}(E,n\le 1000) & :E 
			> E_{\rm cut}.
		\end{array}\right.
\end{equation}
We use $E_{\rm cut} = 64$~eV for \fenineplus and $64.2$~eV for \fetenplus.
The scaling factors used in this small correction are chosen as 
$\kappa = 1.46$ and $2.00$ for \fenineplus and \fetenplus, respectively,
as discussed in Secs.~\ref{fe9mbrrc} and \ref{fe10mbrrc} above. 
The uncertainties are below $1\%$ and can thus be 
safely ignored in the error budget.
From this modified MBRRC we derive a PRRC 
$\alpha_\mathrm{P,hi}(T)$ by convolving the MBRRC 
with an isotropic Maxwellian electron energy
distribution \citep{Schippers2001}.
The final total DR PRRC $\alpha_\mathrm{P}(T)$ is the sum of 
$\alpha_\mathrm{P,lo}(T)$ and $\alpha_\mathrm{P,hi}(T)$.

\subsection{PRRC Results and Comparison}

\begin{figure}
    \plotone{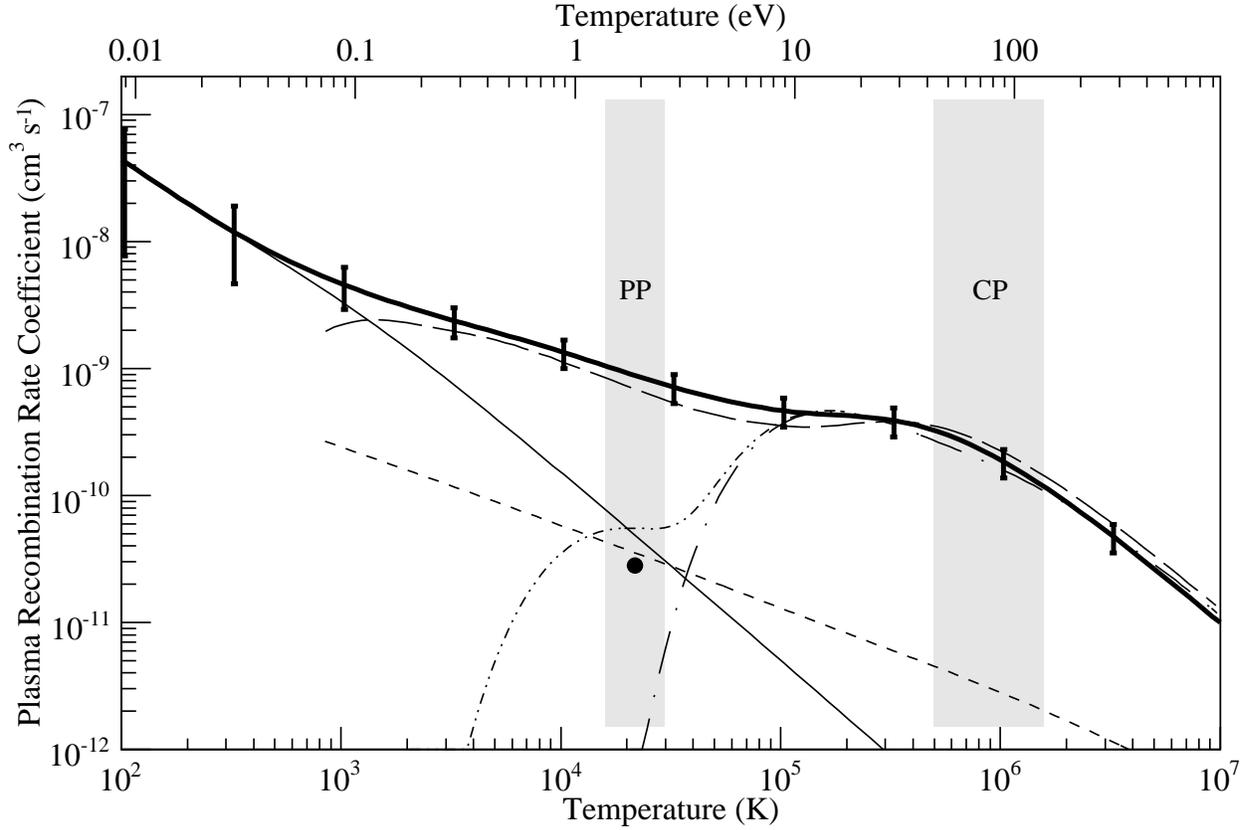}
    \caption{\label{fig:fe9_plasma}
	Comparison of experimental and theoretical \fenineplus PRRC.
    	The thick solid line is the experimental DR PRRC.
	The error bars show the experimental uncertainty at a $90\%$ confidence level.
    	The thin solid line shows the contribution arising solely from the fitted DR 
	resonances below $0.1$~eV, $\alpha_\mathrm{P,lo}(T)$. 
	The long-dashed line is the DR result of \cite{Badnell2006a} and
	the short-dashed line the RR result from the same work.
	The recommended rate coefficient of \cite{ArnaudRaymond1992}
	is shown using a long-dash-dotted curve and its
	modification by \cite{Netzer2004} using a dash-dot-dotted line.
	The filled circle is the estimate made by \cite{Kraemer2004}.
    	The shaded areas indicate the plasma temperatures where the 
	\fenineplus abundance is $\ge 1\%$ in photoionized plasmas
	\citep[PP;][]{Kallman2004} and in collisionally ionized plasmas
	\citep[CP;][]{Bryans2006}. 
    }
\end{figure}

\begin{figure}
    \plotone{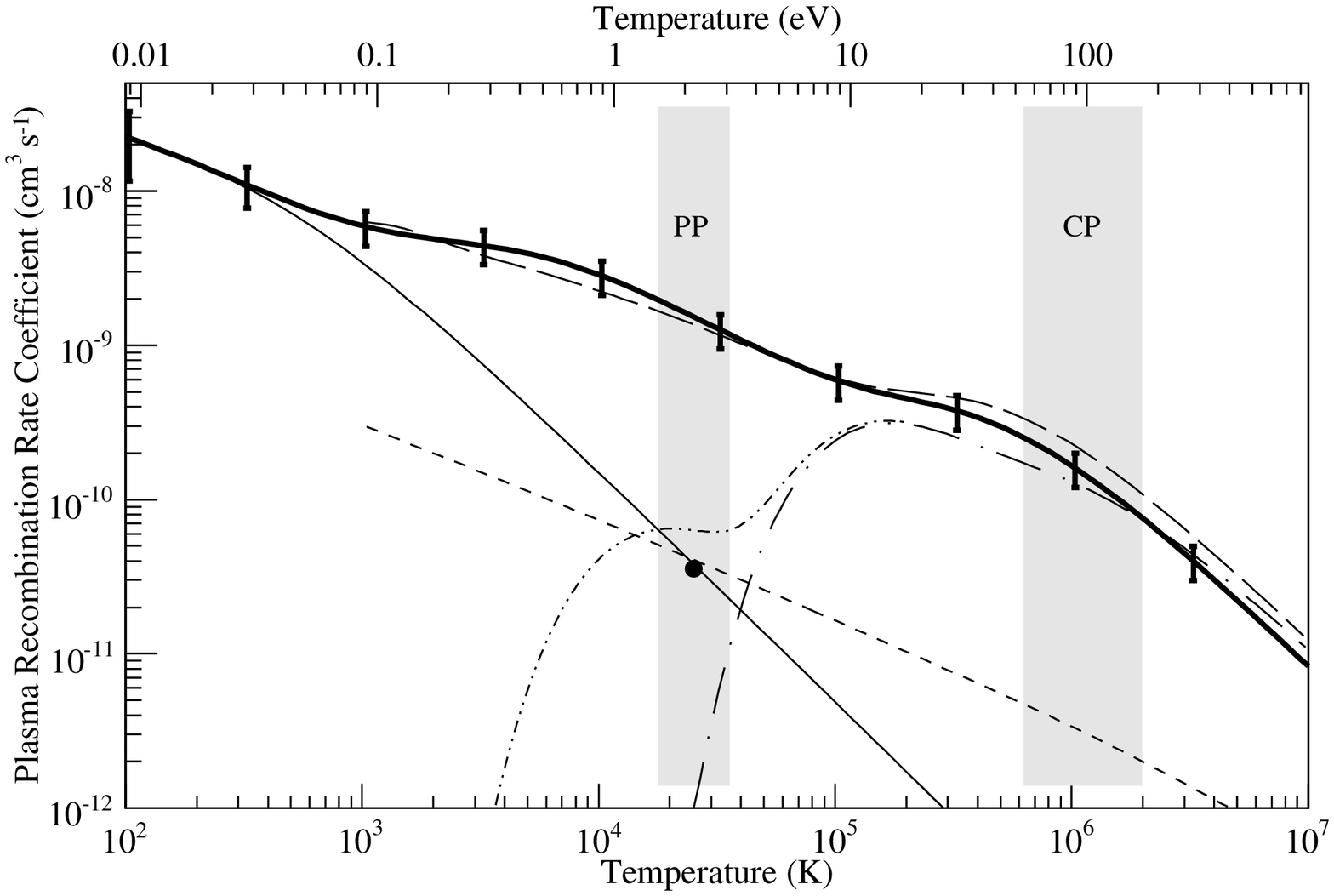}
    \caption{\label{fig:fe10_plasma}
	Same as \figurename~\ref{fig:fe9_plasma} but for \fetenplus.
    }
\end{figure}

Figures~\ref{fig:fe9_plasma} and \ref{fig:fe10_plasma} show the 
experimentally derived DR PRRC (thick solid line) in the temperature range 
from $10^2$--$10^7$~K for \fenineplus and \fetenplus, respectively.
The contributions from DR resonances below $0.1$~eV, $\alpha_{\rm 
P,lo}(T)$,
are shown with a thin solid line and are less than $10\%$ of 
$\alpha_\mathrm{P}(T)$ for both ions at temperatures relevant to their 
formation in photoionized plasmas and negligible for collisionally ionized 
plasmas.
For \fenineplus, the size of the error bars is $82\%$ at $10^2$~K, $67\%$ 
at $3.3\times 10^2$~K,
$36\%$ at $10^3$~K,
$27\%$ at $3.3\times 10^3$~K, and
$25\%$ from $10^4$~K and up.
For \fetenplus the uncertainty is
$47\%$ at $10^2$~K,
$29\%$ at $3.3 \times 10^2$~K, and $25\%$ from $10^3$~K on.
The temperature ranges where the fractional abundance of each ion is $\ge 
1\%$ of the total Fe abundance in photoionized plasmas (PP) and in 
collisionally ionized plasmas (CP) are indicated as grey shaded areas 
\citep{Kallman2004, Bryans2006}.

Also shown in Figs.~\ref{fig:fe9_plasma} and \ref{fig:fe10_plasma} are
the recommended DR rate coefficients of \cite{ArnaudRaymond1992} plotted 
using a long dash-dotted line.
They severely underestimate the DR PRRC of both ions for plasma
temperatures below $10^{5}$~K,
which are of particular importance for photoionized plasmas.
Similar behavior was observed for other M-shell iron ions 
\citep{Schmidt2006, Lukic2007, Schmidt2008}.
As discussed in Sec.~\ref{theory}, this discrepancy arises mostly
from the fact that the underlying theoretical
calculations were carried out for high-temperature plasmas.
\cite{Netzer2004} attempted to modify the results of 
\cite{ArnaudRaymond1992}  in an ad-hoc manner, but as is clearly seen in 
Figs.~\ref{fig:fe9_plasma} and \ref{fig:fe10_plasma} (thin dash-double-dotted 
line) his guess was still too low.
\cite{Kraemer2004} presented a guess for the plasma rate coefficients 
for Fe M-shell ions based on results from L-shell C, N, O, and Fe ions.
Their result (solid circle) are also too low.

The most recent theoretical DR rate coefficients by \cite{Badnell2006b} 
are shown in Figs.~\ref{fig:fe9_plasma} and \ref{fig:fe10_plasma} by the
long-dashed line.
As discussed in Sec.~\ref{theory}, the calculations carried out in the 
frame of the present work result in PRRC insignificantly different from 
those of \cite{Badnell2006b}.
The RR rate coefficients from the same work are
drawn with a short-dashed line. These results were specified only for
the temperature range of $(10^{1} - 10^{7}) \times z^2$~K, where $z$ is 
the parent ion charge state.
These recent calculations are in reasonable agreement with our
experimentally derived PRRCs, though some discrepancies with the 
experimental data can still be observed.
For \fenineplus theory differs from experiment
by factors of $0.80$--$0.76$ for photoionized plasma and
by $1.08$--$1.22$ for collisionally ionized plasma.
Here and below in each pair of numbers quoted the first corresponds to the 
lower temperature limit and the second to the higher temperature limit of 
the cited formation zone.
These differences are within the experimental uncertainties.
For \fetenplus we find factors of $0.85$--$0.93$ for photoionized plasmas 
and $1.33$--$1.42$ for collisionally ionzed plasmas. The deviation at high 
temperatures arises mostly from the larger theoretical resonance strengths 
relative to the experiment
towards the series limits (see Sec.~\ref{fe10mbrrc}).
For both ions the reasonable agreement between experimental and 
theoretical DR plasma rate coefficients is surprising considering the poor
agreement between the measured and predicted MBRRC resonance structure.

\subsection{Parameterization of Plasma Rate-Coefficients}
\label{prrc_fitting}

We have parameterized our experimentally derived DR plasma recombination
rate coefficients $\alpha_{\rm P}(T)$ by fitting the function
\begin{equation}
  \label{eqn:plasmafit_DR}
  \alpha^{\mathrm{fit}}_{\mathrm{P}}\,(T) = T^{-3/2}\, \sum\limits_i c_i 
  \exp(-E_i/T)
\end{equation}
to $\alpha_{\rm P}(T)$.
The fitting results for the free parameters $c_i$ and $E_i$ are given
in Table~\ref{table:plasmafit} for both ions. 
These fits are valid for the energy range of $10^2$ to $10^7$~K,
the residuals
$R = [\alpha_{\rm P}(T) - \alpha^{\rm fit}_{\rm P}(T)] / \alpha_{\rm P}(T)$,
are less than $\pm 3\%$ for temperatures between $10^{2}$ and $10^{5}$~K 
and below $\pm 1\%$ above.

\section{Summary}
\label{conclusion}

We have measured the MBRRC for \fenineplus and \fetenplus at the TSR
heavy-ion storage-ring.
The MCBP theory of \cite{Badnell2006b} and our experimental DR 
results for \fenineplus (\fetenplus) are in poor agreement with regards to
the resonance structure for energies below $48$ ($35$) eV.
Better agreement is found with the integrated resonance strength over this
energy range. 
Above $48$ ($35$) eV we find good agreement for the resonance structure but 
poor agreement for the integrated resonance strengths with theory being 
larger than experiment.
Our results suggest that more specific theoretical
quantities such as DR-generated spectral line excitations could be
considerably uncertain.
The origin of these deviations, which have already been observed in 
previous Fe M-shell studies \citep{Schmidt2006, Lukic2007, Schmidt2008},
remains unknown.

From our measured data, we extracted the DR contributions and transformed
them into a DR PRRC for each ion over a temperature range of
$10^2$--$10^7$~K.
This range includes the temperatures where each ion is abundant in
both photoionized and collisionally ionized plasmas.
As expected, the recommended DR data of \cite{ArnaudRaymond1992} 
underestimates the actual DR PRRC by several orders of magnitude at 
temperatures relevant for photoionized plasmas.
This continues a trend from our earlier measurements for other Fe M-shell
ions \citep{Schmidt2006, Lukic2007, Schmidt2008}.
The ad-hoc estimates by \cite{Netzer2004} and 
\cite{Kraemer2004} turn out to be still too low.
Despite the differences between experiment and state-of-the-art MCBP theory,
the MCBP PRRC for each ion agrees surprisingly well with the experimentally
derived value.
The varying ratios of theoretical to experimental integrated resonance
strengths tend to cancel each other due to the temperature-averaging of 
the PRRC, producing reasonable agreement over a wide range of temperatures.
For photoionized gas, the results for both ions agree within the
uncertainty limits.
For collisionally ionized gas, the theoretical PRRC results have 
a tendency to be somewhat larger than the experimentally derived results.

\acknowledgments
We appreciate the efficient support by the accelerator and \tsr group
during the beam time. 
M. L. thanks Elmar Tr\"abert for the discussion of metastable populations.
This work was supported by the German federal research funding agency DFG 
under contract no.\ Schi 387/5. M.L., D.V.L.\ and D.W.S.\ were supported
in part by the NASA Astronomy and Physics Research and Analysis program,
and the Solar and Heliospheric Physics program.

\appendix
\section{List of abbreviations}
\begin{description}
\item[AGN] active gallactic nucleus
\item[CP] collisionally ionized plasma
\item[DR] dielectronic recombination
\item[MCBP] multi-configuration Breit-Pauli
\item[MBRRC] merged-beams recombination rate coefficient
\item[PP] photoionized plasma
\item[PRRC] plasma rate coefficient
\item[RR] radiative recombination
\item[UTA] unresolved transition array
\end{description}



\begin{deluxetable}{lr}
\tablecaption{\label{table:fe9energylist}
	      Energy levels of \fenineplus relative to the 
	      $3s^2\,3p^5\ [^2{\rm P}^o_{3/2}]$ ground level
	      \citep{AtomicSpectraDatabase} for excitations within the 
	      M-shell.
	      We are unaware of any published data for any of the excited 
	      core levels not listed here.}
\tablehead{\colhead{Level} & \colhead{Energy (eV)}}
\tablewidth{0pt}
\startdata
    $3s^2\,3p^5\ [^2{P}^o_{1/2}]$			& \phn1.94446 \\

    $3s\,3p^6\ [^2{S}_{1/2}]$ 				& 35.8623\phn \\

    $3s^2\,3p^4\,(^3{P})\,3d\ [^4{D}_{7/2}]$ 		& 48.1938\phn \\
    $3s^2\,3p^4\,(^3{P})\,3d\ [^4{D}_{5/2}]$ 		& 48.1938\phn \\
    $3s^2\,3p^4\,(^3{P})\,3d\ [^4{D}_{3/2}]$ 		& 48.3600\phn \\
    $3s^2\,3p^4\,(^3{P})\,3d\ [^4{D}_{1/2}]$ 		& 48.5466\phn \\

    $3s^2\,3p^4\,(^3{P})\,3d\ [^4{F}_{9/2}]$ 		& 51.7824\phn \\
    $3s^2\,3p^4\,(^3{P})\,3d\ [^4{F}_{7/2}]$ 		& 52.4199\phn \\
    $3s^2\,3p^4\,(^3{P})\,3d\ [^4{F}_{5/2}]$ 		& 52.9119\phn \\
    $3s^2\,3p^4\,(^3{P})\,3d\ [^4{F}_{3/2}]$ 		& 53.1022\phn \\

    $3s^2\,3p^4\,(^1{D})\,3d\ [^2{P}_{3/2}]$		& 53.5522\phn \\

    $3s^2\,3p^4\,(^1{D})\,3d\ [^2{D}_{3/2}]$ 		& 53.8853\phn \\

    $3s^2\,3p^4\,(^3{P})\,3d\ [^4{P}_{1/2}]$		& 53.9083\phn \\
    $3s^2\,3p^4\,(^3{P})\,3d\ [^4{P}_{5/2}]$		& 54.7828\phn \\

    $3s^2\,3p^4\,(^3{P})\,3d\ [^2{F}_{7/2}]$ 		& 54.6572\phn \\
    $3s^2\,3p^4\,(^3{P})\,3d\ [^2{F}_{5/2}]$ 		& 56.1314\phn\tablenotemark{a} \\

    $3s^2\,3p^4\,(^1{D})\,3d\ [^2{G}_{9/2}]$ 		& 55.8860\phn \\
    $3s^2\,3p^4\,(^1{D})\,3d\ [^2{G}_{7/2}]$ 		& 55.9273\phn \\

    $3s^2\,3p^4\,(^1{D})\,3d\ [^2{F}_{5/2}]$ 		& 59.1031\phn\tablenotemark{a}\\
    $3s^2\,3p^4\,(^1{D})\,3d\ [^2{F}_{7/2}]$ 		& 60.2542\phn \\

    $3s^2\,3p^4\,(^1{S})\,3d\ [^2{D}_{5/2}]$ 		& \nodata\tablenotemark{b}\\

    $3s^2\,3p^4\,(^1{S})\,3d\ [^2{D}_{3/2}]$ 		& 63.4551\phn \\

    $3s^2\,3p^4\,(^1{D})\,3d\ [^2{S}_{1/2}]$ 		& 67.1844\phn \\

    $3s^2\,3p^4\,(^3{P})\,3d\ [^2{P}_{3/2}]$ 		& 69.9516\phn \\
    $3s^2\,3p^4\,(^3{P})\,3d\ [^2{P}_{1/2}]$ 		& 70.6691\phn \\

    $3s^2\,3p^4\,(^3{P})\,3d\ [^2{D}_{5/2}]$ 		& 71.0372\phn \\
    $3s^2\,3p^4\,(^3{P})\,3d\ [^2{D}_{3/2}]$ 		& 72.6850\phn \\

    $3s\,3p^5\,(^3{P^o})\,3d\ [^4{F}^o_{9/2}]$		& 86.3750\phn \\
    $3s\,3p^5\,(^3{P^o})\,3d\ [^4{F}^o_{7/2}]$		& 86.7259\phn \\
    $3s\,3p^5\,(^3{P^o})\,3d\ [^4{F}^o_{5/2}]$		& 87.1094\phn \\
    $3s\,3p^5\,(^3{P^o})\,3d\ [^4{F}^o_{3/2}]$		& 87.4622\phn \\
\enddata
\tablenotetext{a}{According to \cite{AtomicSpectraDatabase} this level may 
not be real.}
\tablenotetext{b}{Value not specified by \cite{AtomicSpectraDatabase}.}
\end{deluxetable}

\begin{deluxetable}{lr}
\tablecaption{\label{table:fe10energylist}
	       Same as Table~\ref{table:fe9energylist}, but for 
	       \fetenplus relative to the $3s^2\,3p^4\ [^3P_2]$ ground 
	       state.}
\tablehead{\colhead{Level} & \colhead{Energy (eV)}}
\tablewidth{0pt}
\startdata
    $3s^2\,3p^4\ [^3{P}_{1}]$				& \phn1.57062	\\
    $3s^2\,3p^4\ [^3{P}_{0}]$				& \phn1.7745\phn\\

    $3s^2\, 3p^4\ [^1{D}_2]$				& \phn4.67961	\\

    $3s^2\, 3p^4\ [^1{S}_0]$				& 10.0197\phn	\\

    $3s\, 3p^5\ [^3{P}^o_2]$				& 35.1567\phn	\\
    $3s\, 3p^5\ [^3{P}^o_1]$				& 36.3470\phn	\\
    $3s\, 3p^5\ [^3{P}^o_0]$				& 37.0915\phn	\\

    $3s\, 3p^5\ [^1{P}^o_1]$				& 44.8627\phn	\\

    $3s^2\, 3p^3\,(^2{P}^o)\,3d\ [^3{P}^o_2]$		& 61.5073\phn	\\

    $3s^2\, 3p^3\,(^2{D}^o)\,3d\ [^3{S}^o_1]$		& 65.2752\phn	\\

    $3s^2\, 3p^3\,(^2{D}^o)\,3d\ [^3{P}^o_{2}]$		& 65.8716\phn	\\
    $3s^2\, 3p^3\,(^2{D}^o)\,3d\ [^3{P}^o_{1}]$		& 67.1238\phn	\\
    $3s^2\, 3p^3\,(^2{D}^o)\,3d\ [^3{P}^o_{0}]$		& 67.1647\phn	\\

    $3s^2\, 3p^3\,(^2{D}^o)\,3d\ [^1{P}^o_{1}]$		& 66.1394\phn	\\

    $3s^2\, 3p^3\,(^4{S}^o)\,3d\ [^3{D}^o_{3}]$		& 68.7244\phn	\\
    $3s^2\, 3p^3\,(^4{S}^o)\,3d\ [^3{D}^o_{2}]$		& 69.6308\phn	\\
    $3s^2\, 3p^3\,(^4{S}^o)\,3d\ [^3{D}^o_{1}]$		& 70.2222\phn	\\

    $3s^2\, 3p^3\,(^2{D}^o)\,3d\ [^1{D}^o_{2}]$		& 71.7695\phn	\\

    $3s^2\, 3p^3\,(^2{D}^o)\,3d\ [^1{F}^o_{3}]$ 	& 73.6503\phn	\\

    $3s^2\, 3p^3\,(^2{P}^o)\,3d\ [^1{P}^o_{1}]$ 	& 77.2521\phn	\\
\enddata
\end{deluxetable}

\begin{deluxetable}{crrl}
\tablewidth{0pt}
\tablecaption{\label{table:fe9integrals}Integrated DR rate coefficients for \fenineplus. 
		The given errors are purely statistical.}
\tablehead{\colhead{Energy range} & 
	   \colhead{$\int \alpha_{\rm theo} dE$} &
	   \colhead{$\int \alpha_{\rm exp} dE$} &
	   \colhead{$\kappa = \frac{\int \alpha_{\rm theo} dE}{\int \alpha_{\rm exp} dE}$} \\
  	   \colhead{(eV)} & \colhead{(cm$^3$ s$^{-1}$ eV)} & \colhead{(cm$^3$ s$^{-1}$ eV)}}
\startdata
$0.003 - 0.5\phn\phn$	& $1.2087 \times 10^{-9}$ &	$1.25(15) \times 10^{-9}$	& 0.97(12)\\
$0.5 - 2\phd\phn$	& $1.1205 \times 10^{-9}$ &	$1.67(5) \times 10^{-9}$	& 0.67(2) \\
$\phn2 - 10$		& $2.4802 \times 10^{-9}$ &	$3.59(2) \times 10^{-9}$	& 0.691(3) \\
$10 - 36$		& $7.4894 \times 10^{-9}$ &	$9.83(1) \times 10^{-9}$	& 0.7619(8) \\
$\phn\phd36 - 48.7$	& $4.3439 \times 10^{-9}$ &	$6.192(3) \times 10^{-9}$	& 0.7015(3) \\
$48.7 - 66\phd\phn$	& $8.6780 \times 10^{-9}$ &	$5.962(1) \times 10^{-9}$	& 1.4556(3) \\
$66 - 72$		& $1.5871 \times 10^{-8}$ &	$1.1462(3) \times 10^{-8}$	& 1.3847(3) \\
\enddata
\end{deluxetable}

\begin{deluxetable}{crrl}
\tablewidth{0pt}
\tablecaption{\label{table:fe10integrals}Same as 
		Table~\ref{table:fe9integrals} but for \fetenplus.}
\tablehead{\colhead{Energy range} & 
	   \colhead{$\int \alpha_{\rm theo} dE$} &
	   \colhead{$\int \alpha_{\rm exp} dE$} &
	   \colhead{$\kappa = \frac{\int \alpha_{\rm theo} dE}{\int \alpha_{\rm exp} dE}$} \\
  	   \colhead{(eV)} & \colhead{(cm$^3$ s$^{-1}$ eV)} & \colhead{(cm$^3$ s$^{-1}$ eV)}}
\startdata
$0.003 - 0.5\phn\phn$	& $1.7762 \times 10^{-9}$ &	$2.39(13) \times 10^{-9}$	& 0.74(4)\\
$0.5 - 2\phd\phn$	& $2.7179 \times 10^{-9}$ &	$3.83(8) \times 10^{-9}$	& 0.71(2) \\
$2 - 5$			& $2.7658 \times 10^{-9}$ &	$2.18(3) \times 10^{-9}$	& 1.27(2) \\
$\phm{10.}5 - 10.2$	& $2.0254 \times 10^{-9}$ &	$2.04(1) \times 10^{-9}$	& 0.993(5) \\
$10.2 - 37.2$		& $1.0569 \times 10^{-8}$ &	$1.0090(2) \times 10^{-8}$	& 1.0474(2) \\
$37.2 - 55.5$		& $5.5325 \times 10^{-9}$ &	$5.158(2) \times 10^{-9}$	& 1.0726(3) \\
$55.5 - 60.2$		& $1.0341 \times 10^{-8}$ &	$5.1761(3) \times 10^{-9}$	& 1.9979(2) \\
$60.2 - 75\phd\phn$	& $1.0633 \times 10^{-8}$ &	$5.936(3) \times 10^{-9}$	& 1.791(1) \\
\enddata
\end{deluxetable}

\begin{deluxetable}{ccccc}
\tablecaption{Fit parameters $c_i$ (cm$^3$\,s$^{-1}$\,K$^{3/2}$)
		and $E_i$ (K) for the experimental DR PRRC  
		for \fenineplus and \fetenplus using 
		equation~\ref{eqn:plasmafit_DR}.
		Here, the notation $x(y)$ denotes $x\times10^{y}$.
		\label{table:plasmafit}
		}
\tablehead{\colhead{} & \multicolumn{2}{c}{\fenineplus}	& \multicolumn{2}{c}{\fetenplus} \\
 	   \colhead{$i$} & \colhead{$c_i$} & \colhead{$E_i$} & 
 	   		   \colhead{$c_i$} & \colhead{$E_i$} }
\tablewidth{0pt}
\startdata
1	& $6.485(-5)$	& $3.994(1)$		& $6.487(-5)$	& $1.101(2)$	\\
2	& $6.360(-5)$	& $5.621(2)$		& $8.793(-5)$	& $5.654(2)$	\\
3	& $3.720(-4)$	& $1.992(3)$		& $4.939(-4)$	& $1.842(3)$	\\
4	& $1.607(-3)$	& $8.325(3)$		& $3.787(-3)$	& $7.134(3)$	\\
5	& $3.516(-3)$	& $2.757(4)$		& $8.878(-3)$	& $3.085(4)$	\\
6	& $7.326(-3)$	& $7.409(4)$		& $5.325(-2)$	& $1.878(5)$	\\
7	& $2.560(-2)$	& $1.552(5)$		& $2.104(-1)$	& $6.706(5)$	\\
8	& $1.005(-1)$	& $4.388(5)$		& \nodata	& \nodata	\\
9	& $1.942(-1)$	& $7.355(5)$		& \nodata	& \nodata	\\
\enddata
\end{deluxetable}

\end{document}